\documentclass[twocolumn,floatfix]{revtex4-1}
\usepackage[utf8]{inputenc}
\usepackage{amsfonts,amsmath,amssymb,graphicx,epstopdf,verbatim}
\usepackage{placeins}
\usepackage{rotating}
\usepackage{epstopdf}
\usepackage{xcolor}
\usepackage[normalem]{ulem}

\begin{document}

\title{Dynamical hysteresis properties of the driven-dissipative Bose-Hubbard model with a Gutzwiller Monte Carlo approach}
\author{D. Huybrechts and M. Wouters}
\affiliation{TQC, Universiteit Antwerpen, Universiteitsplein 1, B-2610 Antwerpen, Belgium}

\begin{abstract}
    We study the dynamical properties of a driven-dissipative Bose-Hubbard model in the strongly interacting regime through a quantum trajectory approach with a cluster-Gutzwiller Ansatz for the wave function. This allows us to take classical and quantum correlations into account. By studying the dynamical hysteresis surface that arises by sweeping through the coherent driving strength we show that the phase diagram for this system is in qualitative correspondence with the Gutzwiller mean-field result. However, quantitative differences are present and the inclusion of classical and quantum correlations causes a significant shift of the critical parameters. Additionally, we show that approximation techniques relying on a unimodal distribution such as the mean field and $1/z$ expansion drastically underestimate the particle number fluctuations. Finally, we show that a proposed mapping of the driven-dissipative many-body Bose-Hubbard model onto a single driven-dissipative Kerr model is not accurate for parameters in the hysteresis regime.
\end{abstract}

\date{\today}

\maketitle

\section{Introduction}
Driven-dissipative systems have been subject of intense study in recent years thanks to the increased availability of various experimental platforms and their possible applications in quantum simulation \cite{Hartmannreview17, NohReview17, SchmidtADP13}. Technological and experimental advances have lead to a plethora of experimental realizations of driven-dissipative systems, such as circuit QED \cite{FitzpatrickPRX17, SchmidtADP13, Carusotto_RMP_2013_quantum_fluids_light, NohReview17, Hartmannreview17, CarusottoNat2020, HouckNatPhys12, FinkPRX17, Ma19}, semiconductor microcavities \cite{Abbarchi2013, Lagoudakis2010, Carusotto_RMP_2013_quantum_fluids_light, Goblot2019, DeveaudBOOK, Kavokin, BallariniNano19, RodriguezPRL17, FinkNatPhys18}, Rydberg atoms \cite{Mueller_2012, Leseleuc2018, BernienNAT2017, RoyRMP2017, Browaeys2020} and ultracold atoms in optical lattices \cite{TomitaSA17}.
Naturally, this has caused an increased interest in their theoretical description, with various studies on dissipative phase transitions and critical phenomena \cite{HuybrechtsPRB20, Landa20, LeePRL13, JinPRX16, BiondiPRA17, LeBoitePRA14, BoitePRL13, VicentiniPRA18, CasteelsPRA16, HuybrechtsPRA19, CasteelsPRA18}. 
In driven-dissipative systems, the steady state is determined by the competition between the coherent Hamiltonian and dissipative (usually Lindbladian) dynamics. Therefore, their properties can deviate drastically from the thermal equilibrium states of the Hamiltonian. Moreover, due to the absence of a free energy for these driven-dissipative systems the powerful framework of statistical physics cannot be applied, making the description of open quantum systems a more challenging and often a computationally more demanding task. This brings the need for efficient numerical approximation methods as exact solutions are often numerically  infeasible. \\

Theoretical efforts are still ongoing and have thus far lead to the development of various frameworks applicable in different regimes, among which the (cluster) mean-field methods \cite{LeePRL13, JinPRX16, BiondiPRA17, LeBoitePRA14, BoitePRL13}, Gaussian \cite{ VerstraelenPRA20, VerstraelenPRR20,Alberton20}  and Gutzwiller \cite{HuybrechtsPRA19, CasteelsPRA18} variational ansatzes within the wave function Monte Carlo formulation \cite{DalibardPRL92, MolmerJOSAB93, CarmichaelPRL93}, the truncated Wigner approximation \cite{VicentiniPRA18}, the Keldysh function integral formalism \cite{Sieberer_2016, MaghrebiPRB16}, matrix product states (MPS) and matrix product operators (MPO) \cite{VerstraetePRL04,  CuiPRL15, VerstraeteAiP08, OrusPRB08, SchroderPRB16} and the Corner renormalization method \cite{RotaPRB17, BiellaPRA17, FinazziPRL15}.\\
\\
One of the systems that has been widely investigated in this context is the driven-dissipative Bose-Hubbard model \cite{Foss-FeigPRA17, VicentiniPRA18, CasteelsNJP16, BiondiPRA17, CasteelsPRA17-2, LeBoitePRA14, BiondiNJP17, FinazziPRL15, VicentiniPRA19, BoitePRL13}.
It combines physics of optical bistability \cite{Drummond_1980}, well known from nonlinear optics, with the phenomenon of the Mott insulator \cite{GreinerNature2002}, familiar from condensed matter physics. Being a bosonic system with in principle an unbound Hilbert space on each lattice site,
it is also a notoriously difficult model to solve computationally. Moreover, in the driven-dissipative regime, there is a quite large set of tunable parameters that can be used to drive the system to various regimes.
Therefore theoretical studies for extended lattices usually focus on a regime where certain approximations are valid. In the limit of a large photon number, the semiclassical truncated Wigner approximation can be used \cite{VicentiniPRA18}. The limit of strong interactions was addressed with the Gutzwiller mean field ansatz \cite{BiondiPRA17}, $1/z$-expansion where both fluctuations and multiple modes are taken into account \cite{BiondiNJP17}.  Unfortunately, the latter method experiences difficulties describing the hysteretic region in the Bose-Hubbard model, predicted by Gutzwiller mean-field theory \cite{BiondiPRA17}, due to convergence issues in the self-consistent approach. The Corner renormalization method was used for various interaction regimes \cite{FinazziPRL15}. It was also proposed to limit the theoretical description to a single mode in momentum space \cite{CasteelsPRA17-2}. 

Where most previous works started from the master equation formulation of the driven-dissipative dynamics, we will use here instead the equivalent quantum trajectory formulation \cite{BreuerBookOpen}. We will combine it with the factorized Gutzwiller wave function ansatz, that remains valid in the strongly interacting regime. Our approach allows us to go beyond the Gutzwiller approach for the density matrix \cite{BiondiPRA17}, because the classical correlations in the system can be described accurately. It is intuitive that these classical correlations are important in the bistability region, because due to tunneling most of the time the cavities are all together in the low or the high particle number state. As we will show below, these correlations are missed with a Gutzwiller density matrix approximation, but are captured by our Gutzwiller wave function ansatz.

In finite systems, it is known that the bistability is destroyed by switching between the low and high photon number branches, which results in a smooth average photon number as function of pumping intensity. In the thermodynamic limit however, the switching time is expected to tend to infinity and true bistability to emerge. In order to access the steady-state predictions in the thermodynamic limit and for long times, we follow Ref. \cite{CasteelsPRA16} and study the dynamical hysteresis and more specifically its scaling as a function of system size and sweeping velocity.

The structure of the paper is as follows. In section \ref{sec:model} we introduce the driven-dissipative Bose Hubbard model with single photon losses. Then we introduce the used methods for the simulation of the dynamics in section \ref{sec:method}, we elaborate on the quantum trajectory method (also known as the wave function Monte Carlo method) and the used wave function Ansatz. Additionally, we discuss the bistable regime and its dynamical properties. In section \ref{sec:dynhyst} we perform a study of the dynamical hysteresis, which includes its properties in the steady-state limit and a study of the compressibility and the correlation functions. Thereafter, we discuss the validity of a mapping of the driven-dissipative lattice Bose Hubbard model onto a single cavity. Finally, we summarize our results in section \ref{sec:conclusions}.

\section{The Model}\label{sec:model}
The model Hamiltonian of the Bose-Hubbard model with nearest-neighbour hopping and a local Kerr non-linearity is given, in the frame rotating with the drive frequency, by
\begin{equation}\label{Ham}
    \begin{split}
        \hat{H} =& \sum_i \left(-\Delta\hat{a}_i^\dagger\hat{a}_i + \frac{U}{2}\hat{a}_i^\dagger\hat{a}_i^\dagger\hat{a}_i\hat{a}_i + F\left(\hat{a}_i + \hat{a}_i^\dagger\right)\right)\\
        & -\frac{J}{z}\sum_{\langle i,j\rangle}\left( \hat{a}_i^\dagger\hat{a}_j + \hat{a}_j^\dagger\hat{a}_i\right),
    \end{split}
\end{equation}
with $\Delta$ the laser detuning from the cavity frequency, $U$ the Kerr non-linearity, $F$ the pumping strength of the coherent drive, $J$ the hopping amplitude, $z$ the number of nearest neighbours and $\sum_{\langle i,j\rangle}$ a sum over all these nearest-neighbour pairs.\\
The dynamics of the driven-dissipative Bose Hubbard model are then governed by a Lindblad master equation \cite{BreuerBookOpen} describing the time evolution of the density matrix
\begin{equation}
    \frac{d\hat{\rho}}{dt} = -i\left[\hat{H}, \hat{\rho}\right] + \frac{\gamma}{2} \sum_i \left(2\hat{a}_i\hat{\rho}\hat{a}_i^\dagger - \hat{\rho}\hat{a}_i^\dagger\hat{a}_i - \hat{a}_i^\dagger\hat{a}_i\hat{\rho}\right),
    \label{eq:master}
\end{equation}
with  $\gamma$ the dissipation rate and the annihilation operator $\hat{a}_j$ being the jump (Lindblad) operator, i.e. the loss of a photon. Exactly solving this equation is numerically infeasible already for a small number of cavities and we need to resort to approximate methods. In this work we will use the quantum trajectory framework combined with a Gutzwiller wave function ansatz.

\section{The Method}\label{sec:method}
\subsection{Quantum trajectories}
The quantum trajectory method, also known as the wave function Monte Carlo, allows one to calculate the systems dynamics on the level of the wave function through stochastic sampling \cite{DalibardPRL92, MolmerJOSAB93, CarmichaelPRL93, BreuerBookOpen}. One relies on a so-called unravelling of the Master equation in individual trajectories. It is worth noting this unravelling is by no means unique and the chosen unravelling scheme can influence the dynamics of a single trajectory. It is only after averaging infinitely many realisations that the dynamics are equivalent to the Lindblad master equation, irrespective of the unraveling scheme. In this work we opt for the photon-counting unravelling \cite{BreuerBookOpen, DalibardPRL92}. The time evolution of a quantum trajectory is given by a deterministic time evolution with an effective, non-hermitian Hamiltonian and a stochastic jump process. The deterministic part of the time evolution of the wave function $\psi$ is given by
\begin{equation}\label{qt_tev}
    \psi\left(t\right) = \frac{\exp{\left(-iHt\right)}\Bar{\psi}}{\left\| \exp{\left(-iHt\right)}\Bar{\psi} \right\|},
\end{equation}
with $\Bar{\psi}$ an initial (normalized) wave function. The non-Hermitian Hamiltonian $H$ is given by
\begin{equation}
    H = \hat{H} - i\frac{\gamma}{2}\sum_i\hat{a}_i^\dagger\hat{a}_i,
\end{equation}
and does not preserve the norm of $\psi$. The jump process is then given by a (stochastically sampled) quantum jump in the time evolution of the wave function
\begin{equation}\label{eq:jump}
    \psi\rightarrow \frac{\hat{a}_i\psi}{\left\| \hat{a}_i\psi\right\|},
\end{equation}
which can be interpreted as the loss of a photon of the cavity $i$ where the jump takes place. 
The probability for the jump to take place at site $i$ in an infinitesimal time interval $dt$ is given by $p=\gamma dt\langle \psi | a^\dag_i a_i | \psi \rangle$. In our simulations the jumping time is sampled from the waiting time distribution \cite{BreuerBookOpen}.
After the occurrence of a jump the time evolution is again given by Eq. \eqref{qt_tev} until another jump occurs. This continues until a desired simulation time is reached. The strengths of this method lie in the reduction of the Hilbert size, the (usually) modest number of needed trajectories and the ability to gain insight in the single shot systems dynamics. 

\subsection{The wave-function ansatz}
Since the quantum trajectory method simulates the time evolution of the wave function, one works in a smaller Hilbert space than when simulating the Lindblad master equation, where the Hilbert space is the square of the wave-functions Hilbert space. Even though this is already a drastic decrease, the Hilbert space size still increases exponentially with system size. One is thus restricted to very small systems or needs to resort to a wave-function ansatz. In this work we will use a cluster-Gutzwiller ansatz for the wave function \cite{HuybrechtsPRA19}, which allows us to take classical correlations and short-range quantum correlations into account. The wave function ansatz is given by
\begin{equation}
    \Psi_{GW}\left(\left\{ \mathcal{C}\right\}\right) = \prod_\mathcal{C}\psi_\mathcal{C},
\end{equation}
where $\mathcal{C}$ denotes the various clusters present in the lattice of cavities. A cluster is given by various cavities of which the collective state is treated in the Kronecker product of their individual Hilbert spaces. When the cluster size is limited to one cavity, only classical correlations are included, since it is restricted to the Hilbert space of an individual cavity and no quantum correlations are present. In this special case the wave-function ansatz is given by the well known Gutzwiller wave-function ansatz. Unless specified otherwise this will be the ansatz used throughout the rest of the work.\\ 

In the mean-field approximation the time evolution of $\Psi_{GW}$ according to \eqref{qt_tev} is determined by the cluster mean-field Hamiltonian
\begin{equation}
    H_\mathcal{C} = \hat{H}_\mathcal{C} + \hat{H}_{\mathcal{B}\left(\mathcal{C}\right)} -i\frac{\gamma}{2}\sum_{i\in\mathcal{C}}\hat{a}_i^\dagger\hat{a}_i,
    \label{eq:HC}
\end{equation}
where $\hat{H}_\mathcal{C}$ contains all contributions from \eqref{Ham} inside the cluster $\mathcal{C}$ and $\hat{H}_{\mathcal{B}\left(\mathcal{C}\right)}$ contains the mean-field contributions of the nearest-neighbour links across the boundary $\mathcal{B}$ of the cluster $\mathcal{C}$, that is
\begin{equation}
    \hat{H}_{\mathcal{B}\left(\mathcal{C}\right)} = -\frac{J}{z}\sum_{\langle i,j \rangle \vert i\in\mathcal{C}, j \notin\mathcal{C}}\left(\hat{a}_i^\dagger\langle\hat{a}_j\rangle + \hat{a}_i\langle\hat{a}_j^\dagger\rangle \right).
\end{equation}
This results in a system of coupled equations for the time evolution of each of the cluster wave functions $\psi_\mathcal{C}$.
The mean-field Hamiltonian \eqref{eq:HC} implements the deterministic time evolution in the manifold of cluster Gutwiller states according to the time dependent variation principle. Since the dissipation is given by the loss of a particle in a given cavity, the quantum jump does not cause the wave function to leave this variational manifold and Eq. \eqref{eq:jump} is readily applicable.
\begin{figure}
    \centering
    \includegraphics[width=0.45\textwidth]{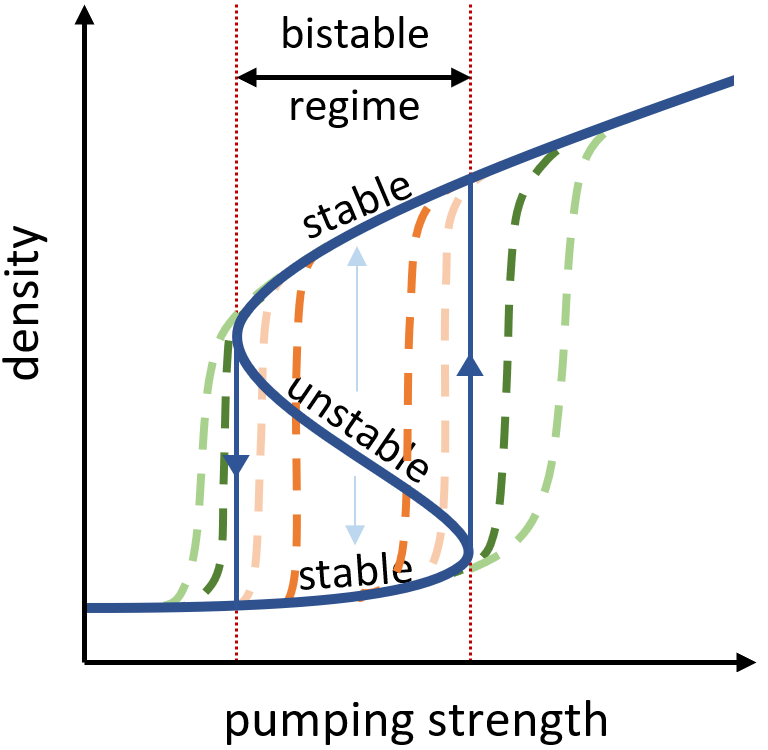}
    \caption{(color online) Example of a bistable region in the driven-dissipative Bose-Hubbard model. There exist two stable branches and one unstable branch, which are the stable and unstable solutions of the mean-field approximation, shown as the blue line. The blue arrows indicates where the system undergoes a first-order phase transition in the steady-state limit, depending on which branch it was on. The two blue arrows mark the boundaries of the bistable regime. The light blue arrow indicates that deviations away from the stable solutions will quickly evolve back to these stable solutions. The dashed light green line shows the dynamical hysteresis curve for a finite sweeping velocity, the dashed dark green line shows the dynamical hysteresis curve for a sweeping velocity which was decreased with respect to the one of the light green dashed line. Similarly, the dashed orange lines show the dynamical hysteresis curves when fluctuations are taken into account, i.e. beyond mean-field, and switching between the branches is possible.}
    \label{fig:bistability}
\end{figure}

\subsection{Bistabile regime and its dynamical properties}\label{sec:model:bistable}

The mean-field equation of the driven-dissipative Bose Hubbard model, consisting of using a Gutzwiller factorized ansatz for the density matrix to solve the master equation \eqref{eq:master}, predicts bistable behaviour for certain parameter regimes \cite{BiondiPRA17}.  The bistable region is a region where the system has two stable states and one unstable state, as shown by the blue line on Fig. \ref{fig:bistability}. Deviations from these stable states quickly cause the system to evolve back to one of these two stable solutions. As such, when the system resides on one of the two stable branches it also stays on this branch. Say the system resides on the bottom stable branch of Fig. \ref{fig:bistability} and the pumping strength is increased across the right boundary of the bistable regime, then the system only has one stable solution, being the top branch, instead of two. The system switches to this new state through a jump, marking a first-order phase transition. If one then decreases the pumping strength the system does not jump back to the lower branch but instead stays on the stable top branch. That is, until the pumping strength crosses the left boundary of the bistable regime and jumps back to the lower stable branch, again marking a first-order phase transition. This is called a hysteretic phase transition.

The above is true for an infinitely slow sweep through the pumping strength which gives information on the steady-state properties. However, instead of using an infinitely slow sweeping velocity to study the steady-state properties one can also study dynamical properties by resorting to a finite sweeping velocity. The faster one sweeps, the less time the system has to adapt to the new pumping strength. This allows the system to reside longer on its previous stable branch before it jumps to its new stable branch. This behaviour is shown on Fig. \ref{fig:bistability} where the dashed dark green line shows the behaviour for a finite sweeping velocity. The dashed light green line shows what is expected to happen when one increases the sweeping velocity with respect to the one of the dashed dark green line. 

Another effect is observed when one allows fluctuations in the system. When these fluctuations are large enough it becomes possible for the system to jump to the other stable branch. As shown on Fig. \ref{fig:bistability} this happens most easily close to the boundaries of the hysteresis curve, since small fluctuations can already drive the system's state into a region where it quickly converges to the other stable state. This behavior is shown as the dashed dark and light orange lines, where the sweeping velocity is faster in the latter one. When the sweeping velocity decreases the system is able to jump to the top (bottom) branch for lower (larger) values of the pumping strength since the system has a longer time to switch between the stable states.  This naturally decreases the hysteretic region \cite{CasteelsPRA16}. The influence of fluctuations then leads to the disappearance of the hysteretic region in the long-time limit in finite-size systems. Only in the thermodynamic limit it is possible for the hysteretic region to survive. This follows from the fact  that as the system gets bigger it is harder for fluctuations to make the entire system switch to the other branch.

Both the green and orange dashed lines show dynamical hysteresis curves. By studying them we gain information on the dynamical properties of the system. It is worth noting that it is hard to determine the dynamical hysteresis region, i.e. the pumping strength interval in which the system shows hysteretic behaviour. To avoid the usage of arbitrary definitions we study the hystertic surface defined as \cite{CasteelsPRA16}
\begin{equation}
    S_h = \int_{F_{start}}^{F_{end}}\left(n_{\uparrow}(F) - n_{\downarrow}(F)\right)dF,
\end{equation}
which is the surface between the top ($n_{\uparrow}$) and bottom ($n_{\downarrow}$) branch of the hystertic curve of the single cavity particle number $\langle\hat{n}\rangle = \frac{1}{N}\langle\sum_i\hat{a}_i^\dagger\hat{a}_i\rangle$, with $N$ the number of cavities in the lattice. $F_{start}$ and $F_{end}$ are chosen in such a way that the entire hystertic surface is enclosed.

The behaviour described above can also be found in the well studied and exactly solvable single cavity \cite{Drummond_1980} in the thermodynamic limit where the photon number tends to infinity while the nonlinearity simultaneously tends to zero. The thermodynamic limit is important here, because otherwise quantum fluctuations cause switching between the two branches, washing out the hysteretic behavior. In the Gutzwiller mean-field solution of the Bose-Hubbard model, the thermodynamic limit is implicit because the equations of motion remain unaltered when the number of sites is increased, hence true bistable behavior is always present in this approximation.
When classical and quantum fluctuations are included in the description of a finite-size driven-dissipative Bose-Hubbard model, this system can also switch between both branches. Incorporating them through the trajectory method and the cluster-Gutzwiller wave function ansatz is therefore expected to give drastic changes in the hysteretic surface, especially in the region where the transitions are expected to happen.

\section{Dynamical hysteresis}\label{sec:dynhyst}
In order to quantify the role of fluctuations in the driven-dissipative Bose-Hubbard model, we start by analyzing the dynamical hysteresis surface. If it tends to a zero value in the limit where the sweep time goes to infinity and in the thermodynamic limit of infinite system size, the system has no hysteretic phase transition. If on the other hand, the hysteretic surface tends to a nonzero value when sweeping slower and slower, even when the system size diverges, the system has a true hysteretic phase transition.
In other words, in the thermodynamic limit and for the sweeping velocity tending to zero there are two possibilities:
(i) the hysteresis surface disappears, resulting in a single curve which is the long-time limit average between both stable branches, i.e. there is one steady-state solution and
(ii) the hysteresis surface converges to a finite value, meaning that when the system resides on one of both branches it will never jump to the other one, i.e. there exist two steady-state solutions.

In the following we perform a linear sweep in pumping strength, that is
\begin{equation}
\begin{split}
    F(t) &=  \left(F_{start} + v_st\right)\theta\left(t < \frac{t_s}{2}\right)\\
    &+ \left(F_{end} - v_s\left(t-\frac{t_s}{2}\right)\right)\theta\left(t \geq \frac{t_s}{2}\right),
\end{split}
\end{equation}
with $t_s$ the total sweep time and $v_s = \frac{2\left(F_{end} - F_{start}\right)}{t_s}$ the sweep velocity. 
Throughout this article we work in the following parameter regime, unless specified otherwise: the relation between the Kerr non-linearity and dissipation rate is chosen as $\frac{U}{\gamma} = 20$ and the laser frequency is tuned to the $4$-photon resonance $1 + \frac{2\Delta}{U} = 4$. The coupling parameter is $\frac{J}{U} = 0.5$. In each lattice we impose periodic boundary conditions. Note that in this parameter region a hysteretic region is predicted in the mean-field approximation  \cite{BiondiPRA17}. The hysteresis surface will be calculated over $F/U$ rather than $F$ and the initial state of the system for each trajectory is given by a lattice of unoccupied cavities.

\begin{figure}
    \centering
    \includegraphics[width=0.5\textwidth]{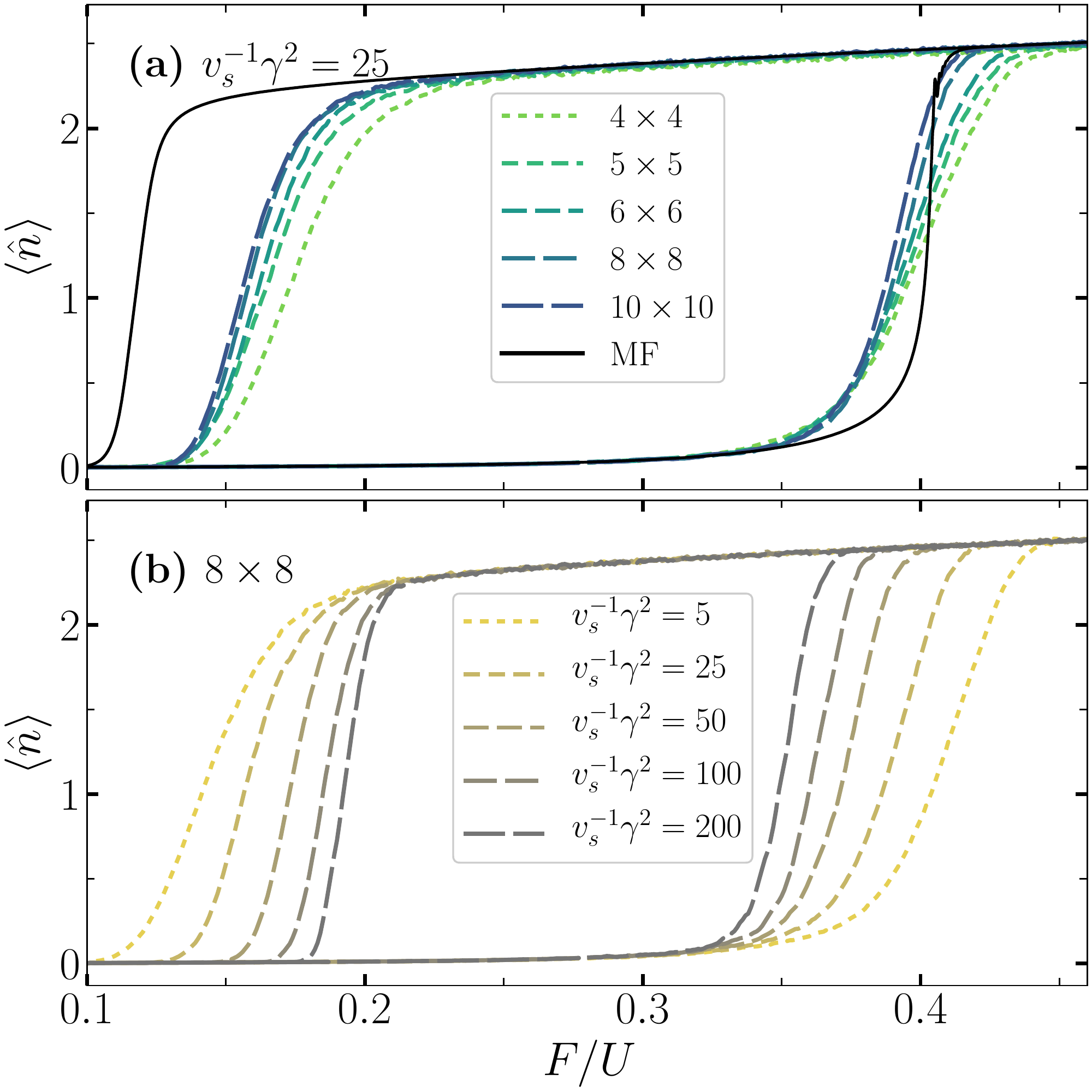}
    \caption{(Color online) \textbf{(a)} Hysteresis curves of the single cavity particle number $\langle\hat{n}\rangle$ for various system sizes (dashed lines) at sweeping velocity $v_s^{-1}\gamma^2 = 25$ and the mean-field (MF) result (full black line) (and $N_{tr}N\approx 10000$, with $N_{tr}$ the number of trajectories, and the cavity Hilbert size cutoff $N_{Max} = 7$). \textbf{(b)} Hysteresis curves of the single cavity particle number $\langle\hat{n}\rangle$ for various sweeping velocities for an $8\times8$ lattice of cavities (and $N_{tr}N\approx 10000$, $N_{Max} = 7$).}
    \label{fig:hystcurves}
\end{figure}

On Fig. \ref{fig:hystcurves} (a) we show hysteretic curves of the average single cavity particle number $\langle\hat{n}\rangle$ for various system sizes with an inverse sweeping velocity of $v_s^{-1}\gamma^2 = 25$. As the system size increases, so does the surface of the hysteresis curves. This coincides with the expectations from the previous section.
Interestingly, it appears that the system does not converge to the mean-field hysteretic surface in the thermodynamic limit. This appears to be due to a large importance of fluctuations on the left side of the bistable region, where a big difference is present for the transition from the top branch to the bottom branch with respect to the mean-field prediction. 
All values for the particle number, not including the ones near the transitions, are however in good agreement with the mean-field results.

On Fig. \ref{fig:hystcurves} (b) the hysteretic surface of a $8\times8$ lattice of cavities is shown for various sweeping velocities. As the sweeping velocities are decreased, i.e. the sweeping time is increased, the hysteretic surface decreases. This is according to the discussion in section \ref{sec:model:bistable}.

For the example of the $8\times8$ lattice of cavities the hysteretic surface is still shrinking as the sweeping velocities are decreased. This continued decrease not only causes the system to move even further away from the mean-field hysteresis surface but also questions whether it will converge to a finite value (ii) or not (i). We perform a more detailed study of the long-time limit in the following section in combination with a finite-size scaling in order to approach the the thermodynamic limit. However, it has to be noted that the previous results give a first indication that through the inclusion of classical correlations the mean-field phase diagram will be qualitatively correct but quantitative differences are expected in the regions where these correlations and fluctuations become increasingly important, i.e. near the phase transition.

\subsection{Steady-state and thermodynamic limit}

\begin{figure}
    \centering
    \includegraphics[width=0.5\textwidth]{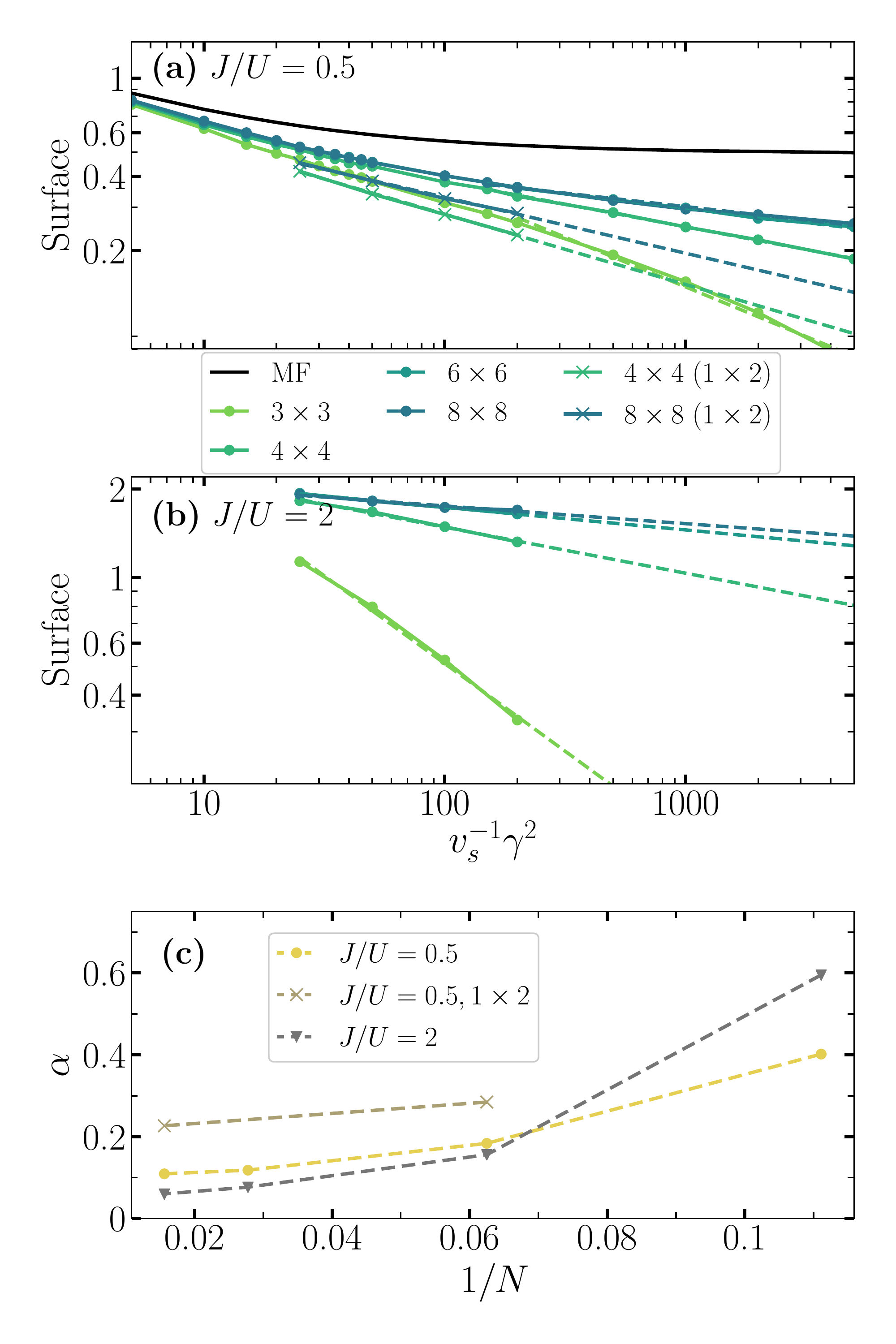}
    \caption{(Color online) \textbf{(a)} [Top Panel] ($J/U = 0.5$) Hysteresis surface for various system sizes at different sweeping velocities for $1\times1$ ($N_{tr}N \approx 10000$, $N_{Max}=7$ and $N_{tr} \approx 100-500$ when $v_s^{-1}\gamma^2 > 200$) and $1\times2$ clusters ($N_{tr}N \approx 5000$, $N_{Max}=7$). The dashed lines are power law fits of the form $S_h = \beta (v_s^{-1}\gamma^2)^{-\alpha}$ in the long time limit.  \textbf{(b)} [Middle Panel] ($J/U = 2$) Hysteresis surface for various system sizes at different sweeping velocities for $1\times1$ clusters ($\text{min}(N_{tr}N) \approx 5000$ and $\text{max}(N_{tr}N) \approx 20000$, $N_{Max}=11$). The dashed lines are again power law fits in the long time limit.  \textbf{(c)} [Bottom Panel] Exponent $\alpha$ of the power law fit $S_h = \beta (v_s^{-1}\gamma^2)^{-\alpha}$ for $1\times1$ clusters at $J/U = 0.5, 2$ and for $1\times2$ at $J/U = 0.5$.}
    \label{fig:twdep}
\end{figure}

Besides studying the long-time limit, i.e. the steady-state, we are also interested in the system's behaviour in the thermodynamic limit. Indeed, phase transitions are only well-defined in the thermodynamic limit and if we wish to compare our results to the steady-state mean-field phase diagram found in Ref. \cite{BiondiPRA17} we need to perform a finite-size scaling to gain access to this infinite-cavity limit. In Fig. \ref{fig:twdep} (a) we show the behaviour of the hysteretic surface of various 2D square lattices as a function of the sweeping velocities for a Gutzwiller ansatz of the wave function, i.e. cluster size one (full lines with circle markers), and a cluster wave function with clusters of size $1\times2$ (full lines with cross markers) with respect to the mean-field result (full black line), at $J/U = 0.5$. Note again that as the sweeping velocity is decreased the size of the hysteresis surface increasingly deviates from the mean-field result. Even as the thermodynamic limit is approached this is not expected to converge, since the results of the $6\times6$ and $8\times 8$ lattices are already practically identical.

By performing a power law fit of the form $S_h = \beta (v_s^{-1})^{-\alpha}$, we can now determine the behaviour in the long-time limit and thus determine whether the hysteretic surface disappears due to the presence of classical (and quantum) fluctuations. Indeed for the observed parameter regime and finite system sizes the extrapolation of the hysteretic surface disappears in the limit of infinite sweeping time. However, it is worth noting that by increasing the system size the exponent of the power law decreases. The question then remains whether the exponent of this power law remains finite in the thermodynamic limit. If it were to become zero we can expect a finite hysteresis surface, i.e. there exists a parameter region where the system shows bistable behaviour in the long-time thermodynamic limit. The scaling of this exponent in the limit of infinite system size is shown as the dashed line with circle markers in Fig. \ref{fig:twdep} (c) . For the observed parameter regime we indeed see a convergence to a non-zero exponent and thus marking the disappearance of the hysteretic region under the influence of classical fluctuations. 

Furthermore, if we include short-range quantum correlations in the form of $1\times 2$ clusters, on top of the classical correlations and on-site quantum correlations this behaviour becomes even more pronounced. This can easily be seen on Fig. \ref{fig:twdep} (a) where we show the hysteretic surface for $1\times 2$ clusters on a $4\times4$ and $8\times 8$ lattice (full lines with cross markers). The inclusion of these short-range quantum correlations causes a further decrease of the hysteresis surface. Moreover, as shown on Fig. \ref{fig:twdep} (c) there is also an increase in the magnitude of the exponent, signalling a faster convergence to a zero hysteretic surface in the thermodynamic limit. It is worth noting that the hysteresis surface exhibits a slow convergence to the power law in the long-time limit, i.e. one needs to simulate trajectories over a long time to obtain the asymptotic behavior. This could lead to (small) corrections on the power law exponents derived here.
Addiationally, we point out that the curves of the hysteresis surface display two power laws, the one studied in the long-time limit on Fig. \ref{fig:twdep} (c), but also the one in the short-time limit. The former is linked to the influence of fluctuations on the system, the latter is linked to the mean field response of the system to the changing pumping strength. Such double power law behavior is also found in experimental and theoretical studies of the single-cavity Bose Hubbard model \cite{CasteelsPRA16, RodriguezPRL17}. 

Our results are in contrast with the mean-field phase diagram from Ref. \cite{BiondiPRA17} where a hysteretic region is predicted at $J/U = 0.5$. Fig. \ref{fig:twdep} indicates that the critical point in the hopping amplitude, where the system switches from a smooth crossover to a hysteretic crossover as the pumping strength is increased, shifts to higher values of $J/U$. This shift is expected to be larger as (longer-range) quantum correlations are included. To investigate more closely the (non)dissapearance of the hysteresis surface as a function of $J/U$ we show Fig. \ref{fig:twdep} (b) where we calculated the hysteretic surface as a function of the sweeping velocity for various system sizes for $J/U = 2$, i.e. roughly an order of magnitude larger than the critical point in the mean-field study (where $J_c/U\approx 0,18$). We indeed see a decrease in steepness of the power law fits and the dashed line with triangle markers on Fig. \ref{fig:twdep} (c) show us that by increasing $J/U$ the exponent has decreased and is approaching zero. From this result we argue that a new (and larger) critical point $J_c/U$ exists where this exponent becomes equal to zero and thus predicts a hysteretic regime. However, a closer study of the region around $J_c/U$ would be needed to pinpoint the exact location of the onset of the hysteretic regime and the nature of the critical point.

We note that in order to obtain a nonzero hysteretic surface one first needs to take the thermodynamic limit before taking the long-time limit. We wish to point out that this observation is in agreement with a study performed in a, related, dissipative spin system in Ref. \cite{Landa20} where an MPO approach was used.

\subsection{The compressibility}

So far we have studied the surface of the hysteresis curve and indicated deviating behaviour with respect to the mean-field method as a result of incorporating fluctuations. We will now explicitly study these particle number fluctuations through the compressibility, that is defined as \cite{BiondiNJP17, BiondiPRA17}
\begin{equation}
\begin{split}\label{eq:compr}
    K = \frac{\langle\hat{N}^2\rangle - \langle\hat{N}\rangle^2}{\langle\hat{N}\rangle} &= 1 - \langle\hat{N}\rangle +\frac{1}{\langle\hat{N}\rangle}\langle\sum_{i,j}\hat{a}_i^\dagger\hat{a}_j^\dagger\hat{a}_i\hat{a}_j\rangle\\
    & = 1+\langle\hat{N}\rangle\big[\sum_{i,j} g_{ij}^{(2)} - 1 \big],
\end{split}
\end{equation}
with $\hat{N} = \sum_i\hat{a}_i^\dagger\hat{a}_i$ and $g_{ij}^{(2)} = \frac{\langle\hat{a}_i^\dagger\hat{a}_j^\dagger\hat{a}_i\hat{a}_j\rangle}{\langle\hat{N}\rangle^2}$. Note that for the Gutzwiller mean-field this reduces to $K = 1 + n\left(g_{ii}^{(2)} - 1\right)$, with $n$ the single site particle number. 

\begin{figure}
    \centering
    \includegraphics[width=0.5\textwidth]{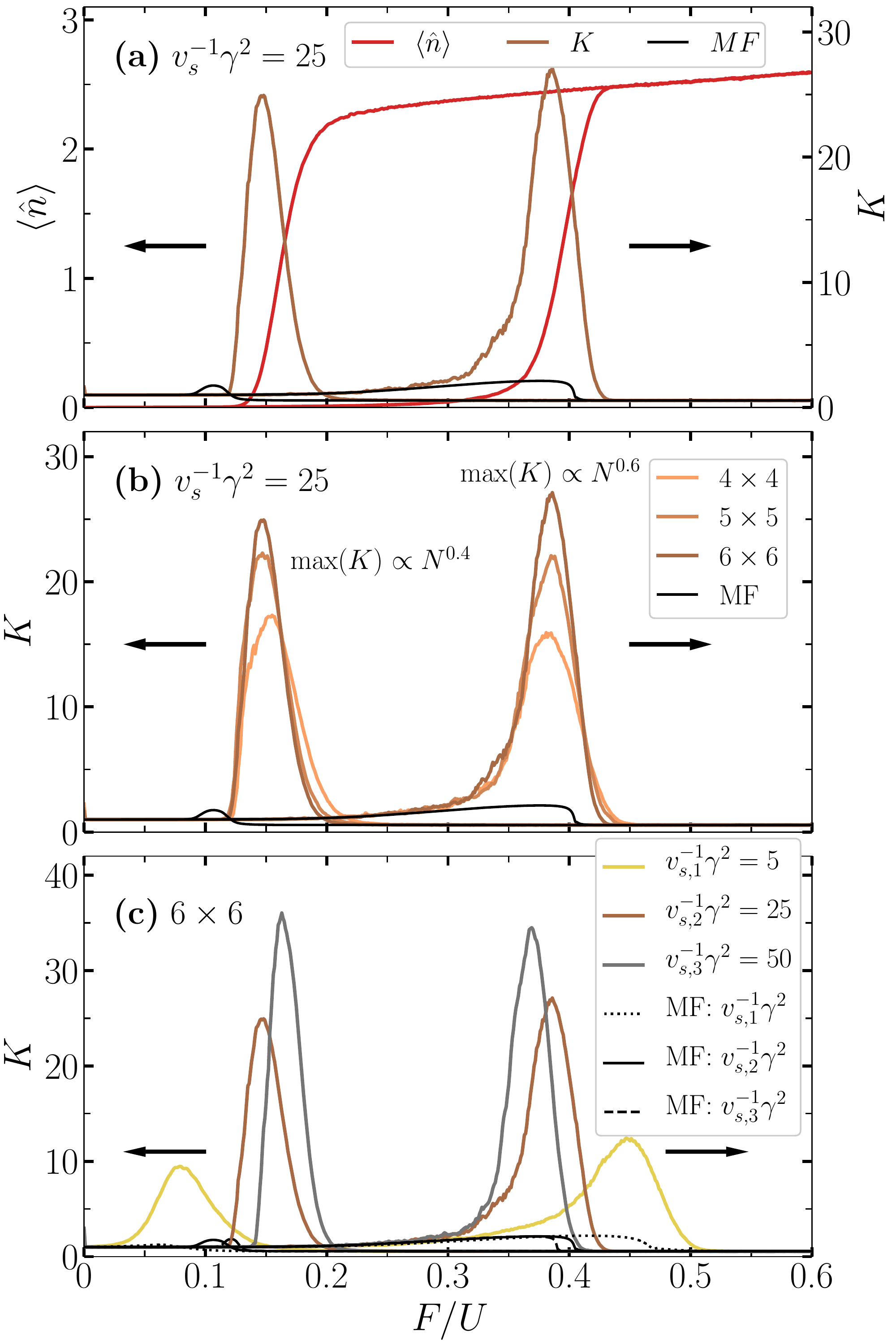}
    \caption{(Color online) \textbf{(a)} Hystersis curve of the particle number per site and compressibility on a $6\times 6$ lattice at sweeping velocity $v_s^{-1}\gamma^2 = 25$ ($N_{tr}N \approx 25000$, $N_{Max}=8$). The mean-field (MF) compressibility is shown as a full black line. \textbf{(b)} Compressibility for various system sizes at a sweeping velocity of $v_s^{-1}\gamma^2 = 25$ ($N_{tr}N \approx 25000$, $N_{Max}=8$). The mean-field (MF) compressibility is shown as a full black line. The power law fits are performed on the maxima of the compressibility of lattices of dimension $5\times5$ to $8\times8$ (not shown). \textbf{(c)} Compressibility of a $6\times 6$ lattice for various for various sweeping velocities ($N_{tr}N \approx 25000$ and $N_{tr}N \approx 40000$ for $v_s^{-1}\gamma^2 = 50$, $N_{Max}=8$). The mean-field (MF) compressibility for the various sweeping velocities are shown as black lines.}
    \label{fig:compr}
\end{figure}

On Fig. \ref{fig:compr} (a) the particle number $\langle\hat{n}
\rangle$ (full red line) and compressibility $K$ (full brown line) of a $6\times 6$ lattice are shown as well as the mean-field compressibility (full black line), both for a sweeping velocity of $v_s^{-1}\gamma^2 = 25$. We confirm that the transitions are accompanied by a presence of high fluctuations which peak at (or close to) the transition. Such a peak can also be observed from the mean-field result albeit with a strikingly lower amount of fluctuations. This discrepancy was expected as the trajectory method allows for jumps between the stable branches, a feature not included in the mean-field method. As our simulations show, this increase in compressibility can be of the order of a magnitude.

Similar to our observations in Fig. \ref{fig:hystcurves} (a), the positions of the maxima of the peaks are also shifted with respect to the mean-field results; at the transition from high to low particle number the mean-field maximum is also found for lower values of $F/U$. The position of the transition from low to high particle number is also similar for both methods. Away from the transition, the results for the compressibility coincide with the mean-field results and deviations only arise close to the transition. 
We have noted before that as the system size increases, it becomes harder for the entire system to switch to the other branch. As a result overall higher fluctuations would be needed for a phase transition to occur in the system. This is indeed what we observe in Fig. \ref{fig:compr} (b) where we show the compressibility for various system sizes at a sweeping velocity of $v_s^{-1}\gamma^2 = 25$. As the system size increases the fluctuations also increase. This is expected since the numerator of \eqref{eq:compr} is quadratic in $\hat{N}$ with respect to the denominator. Through a finite-size scaling we find a sub-extensive power-law scaling with system size of the maxima of both the left and right peaks, of which the exponents are respectively $0,4$ and $0,6$. This sub-extensive scaling is explained by the formation of domains when the system switches to another branch.

Additionally, we can study the behaviour of the compressibility for various sweeping velocities. We show the results of a $6\times 6$ lattice on Fig. \ref{fig:compr} (c). When a fast sweep (full yellow line) is performed, i.e. a high sweeping velocity, the system is driven through the hysteresis region at a very fast pace. The system thus does not have a lot of time to jump between the branches, but it does need to adapt its state very quickly to the new parameter values. This results in fluctuations which are smeared out, explaining the broader, but lower, peaks for high sweeping velocities. This effect decreases as the system gains more time to adapt to the sweeping parameter (full brown and gray line), resulting in narrower and taller peaks. The increased hight of the peaks is due to the increased time the system has to jump between the branches, resulting in more fluctuations. This also means the system will be able to jump to the other branch at smaller (bigger) values for $F/U$ when it is on the bottom (top) branch, which is why the peaks move towards each other as the sweeping velocity is decreased. This is in accordance with the shrinking hysteresis surface observed in the previous section.

The effect of including short-range quantum correlations is shown on Fig. \ref{fig:gij} (a) for $1\times 2$ and $1\times 3$ clusters for a $6\times 6$ lattice at sweeping velocity $v_s^{-1}\gamma^2 = 25$. There is no drastic change in the amount of fluctuations under the influence of the $1\times 2$ clusters, however, they do cause a shift in the position of the transition. When we go one step further in the range of the quantum correlations, i.e. $1\times 3$ clusters, no significant further shift is observed. This is an interesting result as it shows that already for small cluster sizes, the systems properties do not change significantly as longer-range quantum correlations are included, i.e. long-range quantum correlations are expected to be less important.
In other words, by including only short-range quantum correlations, and thus modest computational resources, one can drastically increase the effectiveness of the approximation. 

\subsection{The correlation function}
\begin{figure}
    \centering
    \includegraphics[width=0.5\textwidth]{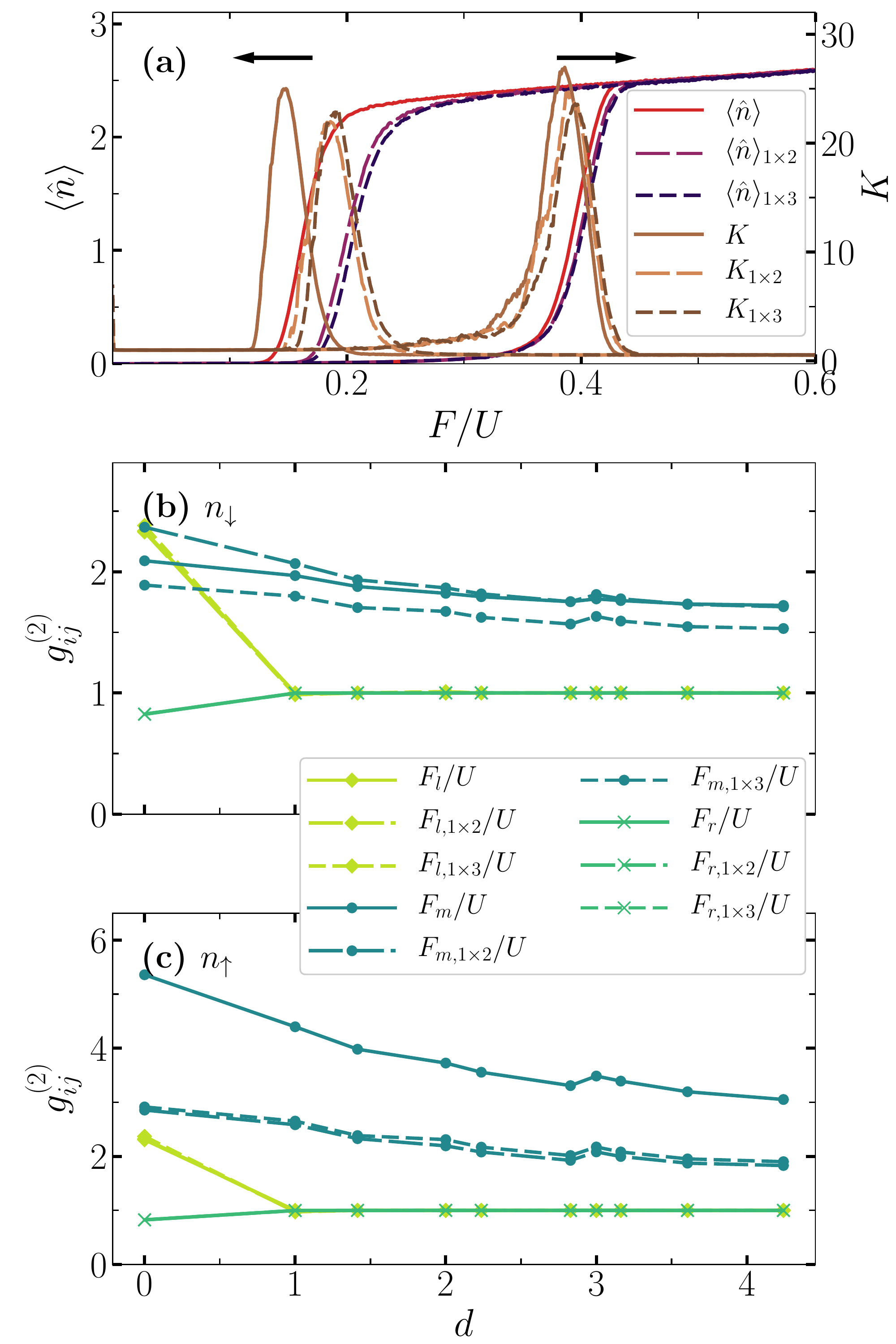}
    \caption{(Color online) \textbf{(a)} Hystersis curve of the particle number per site and compressibility for various cluster sizes on a $6\times 6$ lattice at sweeping velocity $v_s^{-1}\gamma^2 = 25$ ($N_{tr}N \approx 25000$, $N_{Max}=8$ and $N_{Max}=9$ for $1\times3$ clusters)). \textbf{(b)} Correlation function for various distances of the bottom branch for the $6 \times6$ lattice of panel (a) with $F_l/U = 0.05$, $F_r/U = 0.5$ and $F_m/U$ on the respective peaks of the compressibility in panel (a). \textbf{(c)} Correlation function for various distances of the bottom branch for the $6 \times6$ lattice of panel (a). }
    \label{fig:gij}
\end{figure}

The calculation of the compressibility required knowledge of the pair-correlation functions $g_{ij}^{(2)}$. We show the results for the correlation functions as a function of the distance $ d = \vert \textbf{i} - \textbf{j}\vert$ for various parameters of $F/U$ left, right and exactly on the maximum of the compressibility peak of the bottom and top branch on Fig. \ref{fig:gij} (b) and Fig. \ref{fig:gij} (c), respectively. For both branches we note that the system shows no off-site correlations in the regions left and right of the maximum in the compressibility. This explains the success of the mean-field approximation in this region. The on-site bunching in the low-particle number phase ($F_l/U$) and on the maximum is due to the 4-photon resonance. We also find slight on-site antibunching in the high-particle number phase. These results are in accordance with the mean-field theory and a self-consistent expansion in the inverse coordination number of the lattice \cite{BiondiNJP17}. We do find deviations for the off-site correlations at the peak of the compressibility, i.e. near the transition. Our method is able to capture the classical long-range correlations in the lattice, which are expected at the transition, and are missed by mean-field theory. It is due to these correlations that we see a big increase in the compressibility with respect to the mean-field theory.

As noted before, the influence of short-range quantum correlations is most pronounced at the transition from high particle number to low particle number, as can be seen on Fig. \ref{fig:gij} (c) in the dashed lines with circle markers. We note a decrease in correlations with respect to the Gutzwiller wave function ansatz, but long-range correlations are still present. This decrease in correlations is responsible for the slightly lower maximum of the compressibility for the cluster simulations in Fig. \ref{fig:gij} (a). 
\section{Mapping of the Bose-Hubbard model onto a single Kerr cavity}\label{sec:mapping}
\begin{figure}
    \centering
    \includegraphics[width=0.5\textwidth]{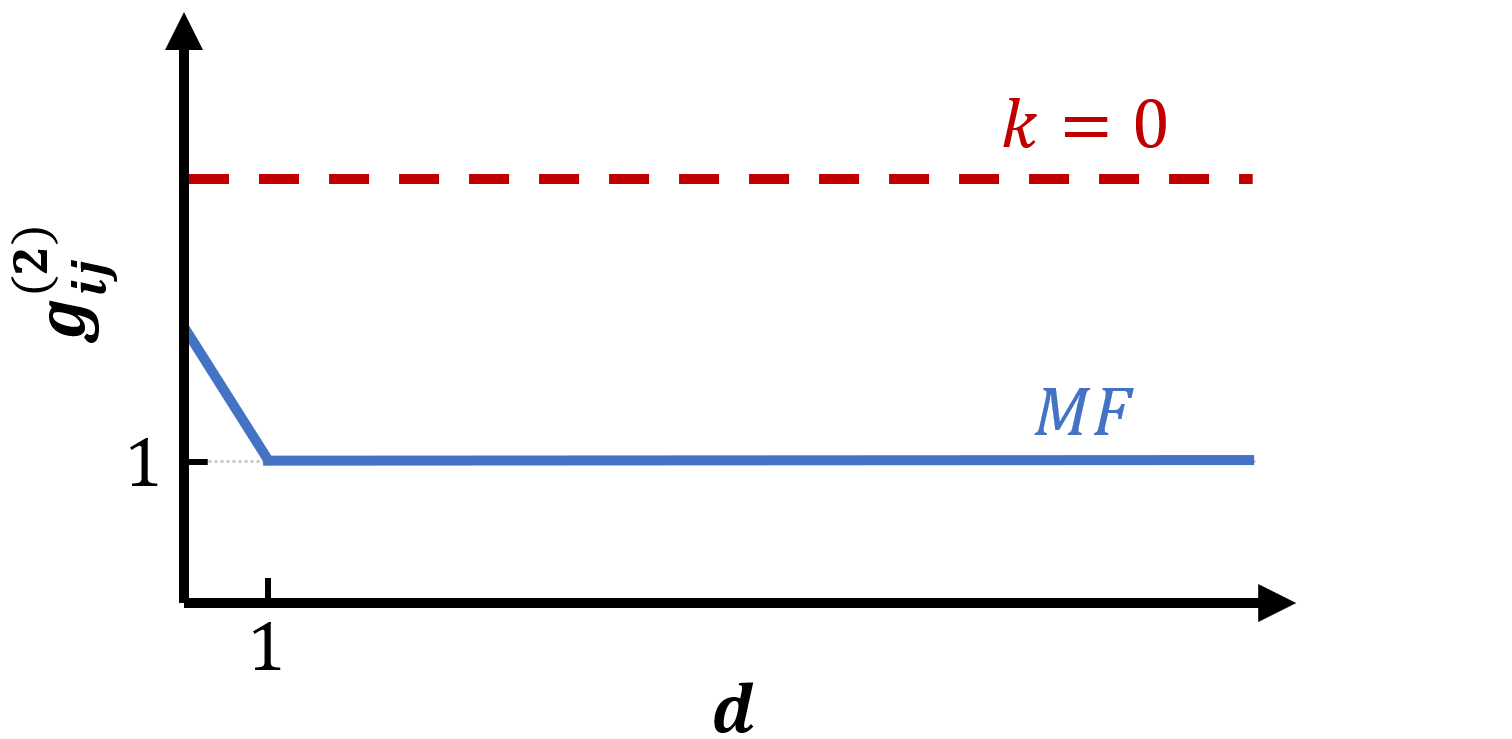}
    \caption{(Color online) A qualitative visualisation of the correlation function for the mean-field method (blue line) and the mapping to the $k = 0$ mode (red dashed line).}
    \label{fig:mapping}
\end{figure}

There exists a mapping of the Bose-Hubbard model with nearest-neighbour hopping \eqref{Ham} on the single Kerr cavity \cite{CasteelsPRA17-2}. The mapping is realized by Fourier transforming the annihilation operator, that is
\begin{equation}
    \hat{a}_i = \frac{1}{\sqrt{N}}\sum_k e^{ikx_i}\hat{a}_k,
\end{equation}
with $x_i$ the spatial coordinates of cavity $i$. By substituting this in \eqref{Ham} one gets the equivalent Hamiltonian
\begin{equation}\label{Hameq}
    \begin{split}
        H &= -\frac{J}{N}\sum_{\langle i,j\rangle}\sum_k 2\cos\left(kd_{ij}\right)\hat{a}_k^\dagger\hat{a}_k -\Delta \sum_k\hat{a}_k^\dagger\hat{a}_k\\
        & +\frac{F}{\sqrt{N}}\sum_i\sum_k\left(e^{-ikx_i}\hat{a}_k^\dagger + e^{ikx_i}\hat{a}_k\right) \\
        &+ \frac{U}{2N}\sum_{k_1,k_2,k_3}\hat{a}_{k_1}^\dagger\hat{a}_{k_2}^\dagger\hat{a}_{k_3}\hat{a}_{k_1 + k_2 - k_3},
    \end{split}
\end{equation}
with $d_{ij} = \vert x_i - x_j\vert$.
The main assumption that is made to map this onto a single Kerr cavity is the following. If one applies a homogeneous drive, only the $k = 0$ mode will be populated. The assumption is then that no nonlinear scatterings will be present and as a result none of the other modes will be populated. This allows all terms $k \neq 0$ in \eqref{Hameq} to be neglected. This results in the following single Kerr cavity Hamiltonian
\begin{equation}\label{Hamsingle}
    H = \omega_0\hat{a}_0^\dagger\hat{a}_0 + F_{eff}\left(\hat{a}_0^\dagger + \hat{a}_0\right) + \frac{U_{eff}}{2}\hat{a}_0^\dagger\hat{a}_0^\dagger\hat{a}_0\hat{a}_0,
\end{equation}
where $\omega_0 = - \Delta - JZ$, $F_{eff} = F\sqrt{N}$ and $U_{eff} = \frac{U}{N}$. This is an interesting mapping because it can be used to study the thermodynamic limit of \eqref{Ham} by tuning $F_{eff}\rightarrow \infty$ and $U_{eff} \rightarrow 0$, keeping the product $U_{eff}F_{eff}^2$ constant. However, neglecting the nonlinear scatterings could make the above mapping invalid in certain parameter regimes.
To study its validity we look again at the correlation function.
In Fig.  \ref{fig:mapping}, we illustrate schematically the behavior of the correlation function in the $k=0$ (red dashed line) and mean-field approximations (blue line).  In the $k=0$ approximation, all cavities are perfectly correlated, such that $g^{(2)}$ is flat. In the mean-field approximation on the other hand, correlations are entirely neglected, such that $g^{(2)}=1$ for $d\neq 0$. 
For pump intensities far from the transition, Fig. \ref{fig:gij} shows that the correlation function is close to one for $i\neq j$. This is consistent with the good agreement with the mean-field theory in this parameter regime. For the correlation function on the upper branch ($F_r$ in Fig. \ref{fig:gij} (c)), the correlation function is very flat, such that it is also compatible with the $k=0$ model. This is however not the case in the low intensity ($F_l$) case. In the transition region, we have already seen that the mean-field approximation breaks down. Unfortunately, the correlation functions at $F_m$ in Fig. \ref{fig:gij} (b) \ref{fig:gij} are also not constant, such that neither the $k=0$ is valid in the transition region. Physically, this is due to the formation of domains of high and low intensity in the switching region, which goes beyond the assumptions of the $k=0$ model.

\section{Conclusions}\label{sec:conclusions}
We have shown that by using the cluster-Gutzwiller Monte Carlo method, and thus the incorporation of classical correlations and short-range quantum correlations, the Gutzwiller mean-field phase diagram is qualitatively correct. The influence of the included fluctuations causes the critical hopping parameter $J_c/U$, indicating where the steep crossover of the system transforms into a first-order hysteretic phase transition, to shift to higher values. From the study of the particle number fluctuations we can conclude that away from the transition the results fall onto the mean-field results, also when close-range quantum correlations are included. This is confirmed by the absence of off-site correlations in these regimes, explaining the success of the mean-field method. At the transition we note a drastic increase in the particle number fluctuations, by an order of magnitude, with respect to the mean-field result. This increase finds it origin in the off-site correlations that are included through our method, and which are absent in mean-field theory. Additionally, a shift in the location of the transition from high to low particle number is also observed. This shift does not change significantly when more short-range quantum correlations are included, signalling that long-range quantum correlations are expected to be less important. We thus show that by a modest increase in computational resources the effectiveness of the approximation can be drastically improved. Finally, we show that care has to be taken when performing the mapping of an extended Bose-Hubbard lattice onto a single Kerr cavity, especially in the proximity of a phase transition, i.e. regions where fluctuations and correlations are important.

\acknowledgements  Discussions with W. Verstraelen, M. Van Regemortel and F. Minganti are greatfully acknowledged. This work is supported by UAntwerpen/DOCPRO/34878. Part of the computational resources and services used in this work were provided by the VSC (Flemish Supercomputer Center), funded by the Research Foundation - Flanders (FWO) and the Flemish Government department EWI.

\bibliography{main.bbl}

\end{document}